%% file: main.tex
\documentclass{article}
\usepackage[utf8]{inputenc}

\usepackage{authblk}
\usepackage{amssymb}
\usepackage{dsfont}
\usepackage{geometry}
\usepackage{amsthm}
\usepackage{amsmath}
\usepackage[hidelinks]{hyperref}
\usepackage{graphicx}
\usepackage{amsmath}
\usepackage{systeme}
\usepackage[round]{natbib}

\counterwithin{figure}{section}
\DeclareMathOperator*{\argmin}{arg\,min}

\title{Through-the-Cycle PD Estimation Under Incomplete Data\\ \Large A Single Risk Factor Approach}

\author{
    Barbara Dömötör\thanks{barbara.domotor@uni-corvinus.hu} { and}
    Ferenc Illés\thanks{ferenc.illes@uni-corvinus.hu}\\
    Institute of Finance, Corvinus University of Budapest
}

\begin{document}
\maketitle
\begin{abstract}
Banks are required to use long-term default probabilities (PDs) of their portfolios when calculating credit risk capital under internal ratings-based (IRB) models. However, the calibration models and historical data typically reflect prevailing market conditions. According to Basel recommendations, averaging annual PDs over a full economic cycle should yield the long-term PD. In practice, the available data are often temporally incomplete -- even for high-risk portfolios. In this paper, we present a method for the simultaneous calibration of long-term PDs across all sub-portfolios, based on the single risk factor model embedded in the Basel framework. The method is suitable even for smaller, budget-constrained institutions, as it relies exclusively on the bank’s own default data. A complete dataset is not required -- not even for any individual sub-portfolio -- as the only prerequisite is the presence of overlapping data before and after the missing values, a mild condition that is typically met in practical situations.\\[10pt]

\noindent
\textit{Keywords:}
\noindent
Vasicek model, through-the-cycle probability of default, credit risk capital, missing data estimation

\end{abstract}

\section{Introduction}
\label{section:intro}
\input{intro.tex}

\section{A framework for PIT and TTC PD}
The calculation of credit risk capital requirements for IRB banks is based on the \textit{Asymptotic Single Risk Factor} (ASRF) model, introduced by Vasicek \citep{vasicek2002distribution}. This model maps the unconditional probability of default (PD) into a conditional PD, expressed as a function of a latent systematic risk factor. From the static version of this model, we can derive the "worst-case" conditional PD that we can expect under extreme bad economic conditions. This allows us to approximate the worst-case PD in terms of the unconditional PD. By extending the model across multiple periods and multiple portfolios, we can reverse this process and estimate the unconditional PD based on a series of conditional PDs.

\subsection{Static Vasicek model}
\label{section:vasicek}
\input{vasicek.tex}

\subsection{Dynamic Vasicek model}
\label{section:model}
\input{model_description.tex}


\section{Model calibration}
\label{section:calibration}
To put it simply, model calibration means solving the constrained optimization problem
\begin{equation}  
\label{eq:optim}
\min\limits_{p_i,\rho_i, f_t} \longrightarrow e\left(\Phi\left(\frac{\Phi^{-1}(p_i) - \sqrt{\rho_i}f_t}{\sqrt{1-\rho_i}}\right), d_{i,j}\right) \qquad \text{s.t.}\ \sum_t f_t = 0,
\end{equation} 
where $e$ is a suitable error function. Though this may seem like a difficult nonlinear optimization problem, it can be easily transformed and solved in three simple steps.
\subsection{Complete linear model}
\label{section:linear_model}
\input{linear1.tex}

\subsection{Incomplete linear model}
\label{section:incomplete_linear_model}
\input{linear2.tex}

\subsection{Non-linear model}
\label{section:nonlinear_model}
\input{nonlinear.tex}

\section{Numerical tests}
\label{section:numeric}
\input{numeric.tex}

\section{Conclusion}
This paper presents a simple, robust method for estimating through-the-cycle (TTC) probability of default for IRB capital calculations using the Vasicek model. The approach relies solely on internally available data, ensuring alignment with the bank’s portfolio. It is mathematically sound, produces stable results even amid portfolio shifts, and remains applicable as long as overlapping subportfolios exist around periods with missing data. The method also allows for the integration of expert judgment or supervisory input related to the current macroeconomic cycle.

\newpage
\bibliography{ref.bib}
\bibliographystyle{plainnat}

\end{document}

%% file: intro.tex
Credit risk is the most significant risk faced by financial institutions, directly affecting both institutional solvency and the resilience of the broader financial system, while accounting for more than 80\% of total regulatory capital requirements \citep{eba2025capital}. Accurate measurement and management of credit risk are essential for effective capital allocation, regulatory compliance, and risk-informed decision-making. 

According to the Basel regulatory framework, banks are required to hold sufficient capital to cover potential credit losses over a one-year horizon at a 99.9\% confidence level. The Basel II framework, introduced in 2007, permits institutions to use internally derived estimates of the probability of default (PD) in the capital calculation \citep{BIS2005}. These PD estimates must reflect long-term, cycle-independent default risk -- commonly referred to as through-the-cycle (TTC) PDs -- designed to capture the average credit risk over an entire economic cycle. This approach aims to reduce procyclicality by smoothing short-term fluctuations and promoting stability in capital requirements. However, credit risk models and historical default rates tend to be risk-sensitive, reflecting prevailing market and borrower conditions, which typically result in point-in-time (PiT) PD estimates (\cite{novotny2016interaction}; \cite{mikolasekproblems}). Accurate PD estimation, however, is critical, as even a one-percentage-point deviation can lead to a 5-15 percentage over- or underestimation of the required capital, depending on the level of default probability -- potentially resulting in a substantial misalignment with regulatory capital requirements.

Several methodological approaches have been developed to achieve long-term, cycle independent PDs. The bottom-up method builds on borrower-specific or segment-level PDs, adjusting observed point-in-time (PiT) estimates to remove cyclical effects using expert judgment, long term historical loss data \citep{aguais2008designing}, macroeconomic indicators (\cite{bellotti2009credit}; \cite{garcia2017approximate}; \cite{engelmann2024simple}), or econometric filtering techniques (\cite{loffler2004anatomy}; \cite{ingolfsson2010cyclical}). Another widely used approach is based on credit rating migration matrices, where transition probabilities between rating classes are estimated over long horizons. These matrices can be smoothed or averaged across different economic periods to isolate the cycle-independent component of credit migration, from which TTC PDs can be inferred (\cite{belkin1998one}; \cite{carlehed2012methodology}; \cite{witzany2022stressing}; \cite{basson2023through}).

The Basel framework suggests calculating the long-term probability of default by averaging PDs that reflect expected default rates under normal business conditions \citep{BIS2005}. While this provides an unbiased estimate, it assumes the availability of consistent data over a full economic cycle—an assumption that is often unrealistic.

To address the problem of data scarcity, \cite{ingolfsson2010cyclical} propose a variable scalar approach, in which the cyclical component is identified with the scalar term of the logit model used for point-in-time (PiT) PD estimation. By adjusting the PiT PDs using this scalar, the method yields consistent through-the-cycle (TTC) estimates, which can be further refined as more data become available. \cite{engelmann2024simple} on the other hand, integrates limited internal data with a macroeconomic model that captures an abstract economic state closely aligned with a bank’s portfolio conditions. A key requirement is the availability of a suitable macroeconomic model that allows the inclusion of information on the full economic cycle into risk parameters estimates even if bank-internal data covers only a small part of an economic cycle. 

This paper builds on the literature addressing data limitations through a Vasicek model-based calibration approach. However, unlike previous studies, we do not model the credit cycle using macroeconomic variables; instead, we derive it directly from observed default data. Under reasonable assumptions, we demonstrate that the calibration process allows for the simultaneous estimation of long-term, unconditional (through-the-cycle) and conditional (point-in-time) default probabilities, as well as the underlying cyclical component. Our method is simple to implement and relies solely on banks’ existing PD models, providing a practical solution for meeting regulatory requirements -- even for smaller institutions.

The paper is organized as follows. Section 2 reviews the static single risk factor model of Vasicek and specifies the theoretical conditions under which its dynamic extension enables the derivation of unconditional (through-the-cycle) probability of default from conditional (point-in-time) values. Section \ref{section:calibration} introduces the proposed calibration methodology utilizing observed loss data, while Section \ref{section:numeric} demonstrates its application on a simulated dataset. Finally, we draw the conclusions.

%% file: vasicek.tex
Let $N$ be a finine set, and for all $i\in N$ let $X_i\in\mathbb{R}$ be a standard normal distributed random variable and let $K\in\mathbb{R}$ be a real number. We say that $i\in N$ is in default when $X_i < K$. The elements of $N$ are called \textit{entities} and they can be loans or firms or any other objects or debtors for which \textit{default} is a meaningful event. $X_i$ can be interpreted as the \textit{state} or quality of entity $i\in N$. We do not specify the meaning of a default, the model simply assumes that it happens when the state of an entity is under the threshold $K$. The probability of the default of each entity is therefore
\begin{equation} 
\label{eq:PD}
\text{\textit{PD}}=\Phi(K),
\end{equation} 
where $\Phi$ is the standard normal (cumulative) distribution function. This will be refered to as \textit{unconditional PD}. The key assumption of the model is how the variables $X_i$ and the defaults of the entities are interrelated. We assume that the entities share a common systematic risk factor, and the defaults are conditionally independent to this factor. 
\begin{equation}  
\label{eq:2factors}
X_i = \sqrt{\rho} F + \sqrt{1-\rho}Z_i,
\end{equation} 
where the systematic factor $F$ and the entity-specific risk factors $Z_i$ are independent standard normal distributed random variables, and $0<\rho<1$ is the correlation between $X_i$ and $X_j$ (if $i\neq j \in N$). To calculate the \textit{conditional} PD to the systematic factor we can simply rearrange the equation to $Z_i$ as follows.
\begin{multline} 
\label{eq:vasicek}
\mathbb{P}\left(X_i < K \ | \ F = f\right)=\mathbb{P}\left(\sqrt{\rho} F + \sqrt{1-\rho}\ Z_i < K \ | \ F = f\right)=\\ = \mathbb{P}\left(Z_i< \frac{K - \sqrt{\rho} f}{\sqrt{1-\rho}}\right)= \Phi\left(\frac{K - \sqrt{\rho}f}{\sqrt{1-\rho}}\right)=\Phi\left(\frac{\Phi^{-1}(PD) - \sqrt{\rho} f}{\sqrt{1-\rho}}\right)
\end{multline}
In the last step we replaced $K$ by $K= \Phi^{-1}(PD)$ based on Equation \eqref{eq:PD}. For a significance level $\alpha\in(0,1)$ the \textit{worst-case-default-rate} on level $\alpha$ ($WCDR(\alpha)$) is the $\alpha$-quantile of the conditional PD distribution. As it is a decreasing function of $f$, we can calculate the WCDR by simply plugging the $(1-\alpha)$-quantile of $F$ into Formula \eqref{eq:vasicek} as follows.
\begin{equation}  
\label{eq:wcdr}
\text{\textit{WCDR}}(\alpha) = \Phi\left(\frac{\Phi^{-1}(PD) - \sqrt{\rho}\Phi^{-1}(1 - \alpha)}{\sqrt{1-\rho}}\right) = \Phi\left(\frac{\Phi^{-1}(PD) + \sqrt{\rho}\Phi^{-1}(\alpha) }{\sqrt{1-\rho}}\right)
\end{equation} 
Current regulation requires to calculate the capital requirement for banks under IRB based on the WCDR for $\alpha = 0.999$, which requires unconditional PD as input.

%% file: model_description.tex
The dynamic Vasicek model is the extention of the static Vasicek model for multiple periods and multiple portfolios. One possible approach is to assume that the systematic factor follows a stochastic process (for example an autoregressive model), and derive analytical results about the distribution of the conditional and unconditional PDs. Such a model is introduced, by \citet{garcia2017approximate}. Here we follow a slightly different approach by considering the unconditional PDs and the factors as (nonrandom) model parameters and the set of conditional PDs as "predictions". Hereby we present the formal description of the model and then we outline how to apply it in practice.

\subsubsection{Formal description}
\label{section:multivasicek}
\begin{itemize}
\item Let $N$ be a finite set, partitioned into $N = \bigcup_{i\in I} N_i$ subsets with some finite index set $I$,
\item let $T$ be a positive integer,
\item for each $i\in I$ let $K_i\in\mathbb{R}$, $\rho_i\in\mathbb{R}$,
\item and for each $1 \le t \le T$ let $f_t\in\mathbb{R}$ be real parameters.
\end{itemize}
Here $T$ denotes time, subsets $N_i$ are called sub-portfolios and consist of entities with the same unconditional PD $p_i = \Phi(K_i)$, correlation parameter $\rho_i$, and for each year $t$ fulfill the conditions of the Vasicek model with systematic factor $f_t$. Based on Formula \eqref{eq:vasicek}, the conditional PD for sub-portfolio $i$ in year $t$ is
\begin{equation}
\label{eq:pit}     
p_{i,t} = \Phi\left(\frac{\Phi^{-1}(p_i) - \sqrt{\rho_i}f_t}{\sqrt{1-\rho_i}}\right).
\end{equation}
\begin{itemize}
\item The \textit{input} of the model is a data set \[d_{i,t}\in(0,1)\] called \textit{empirical default rates} for sub-portfolio $i$ in year $t$, which can be incomplete, meaning that for some $i$ and $t$ pairs $d_{i,t}$ might not be available, in which case we use the notation $d_{i,t}=$\textit{NA} (missing value). 
\item By \textit{calibration} or \textit{fitting} the model, we mean
\begin{itemize}
\item a definition of an error function $r:\mathbb{R}^*\times\mathbb{R}^*\to\mathbb{R}$, where $\mathbb{R}^*$ denotes the set of finite (but arbitrarily long) sequences of real numbers, 
\item a process to find model paramaters $K_i$, $\rho_i$ ($i\in I$) and $f_t$ ($1\le t \le T$) that minimize the error $e(p_{i,t},d_{i, t})$ over $i$ and $t$ pairs for which $d_{i, t}\neq$\textit{NA}.
\end{itemize}
\item The \textit{output} of the model is the optimal $p_i$, $\rho_i$ and $f_t$ model parameters and the entire set of fitted $p_{i,t}$ values, where $p_i$ is interpreted as unconditional PD, and $p_{i,t}$ is interpreted as conditional PD.
\end{itemize}
Calibration methods will be suggested and discussed in Section \ref{section:calibration}. However, let us point out that the model parameters are not unique without further constraints, because the left hand side of Equation \eqref{eq:pit} does not define the parameters on the right hand side. For any parameters set $p_i$, $\rho_i$ and $f_t$  
\[\Phi\left(\Phi^{-1}(p_i)+\lambda\sqrt{\rho_i}\right), \ \rho_i \ \text{ and } \ f_t+\lambda\]
defines the same $p_{i,t}$ values.
This is an identifiability problem. Identifiability is a models ability to learn its parameters from data. The Vasicek model is not identifiable, because it cannot distinguish the effects of the systematic and the individual factors. A parallel shift in $f_t$ can be compensated by an adjustment of the $p_i$ unconditional PDs to get the same conditional PDs. By applying the inverse normal distribution function to both sides of Equation \eqref{eq:pit} and multiplying it by the denominator, we get the
\begin{equation}
\label{eq:y}  
	y_{i, t} = \varphi_{i,t}\sqrt{1-\rho_i} = K_i-\sqrt{\rho_i} f_t,
\end{equation}
linear relationship between transformed parameters 
\begin{equation}
\label{eq:yphi}  
	\varphi_{i, t} = \Phi^{-1}(p_{i,t}) \quad \text{and} \quad y_{i,t}=\varphi_{i,t}\sqrt{1-\rho_i}.
\end{equation}
Let us average $y_{i, j}$ for a given, fix $i$ sub-portfolio. We get that
\[K_i - \sqrt{\rho_i} \frac{\sum_t{f_t}}{T}  = \frac{\sum_j y_{i,t}}{T},\]
which immediately implies that 
\[K_i = \frac{\sum_j y_{i,t}}{T} \Longleftrightarrow \frac{\sum_t{f_ t}}{T} = 0.\]
In words, the time-mean of the transformed conditional PDs is the transformed unconditional PD for \textit{all} sub-portfolios at the same time if and only if the time-mean of the systematic factor is zero. Let us point out that in the original, static Vasicek model the systematic factor has standard normal distribution, so its expected value is zero. If we add the extra constraint
\begin{equation}
\label{eq:f0}
\frac{\sum_t{f_t}}{T} = 0,
\end{equation}
the model becomes identifiable and parameters $p_i$ and $f_t$ are uniquely determined. As the zero mean for $f$ can be interpreted as the representativeness of the data, we can relax this condition. If the time interval $1 \le t \le T$ does not contain enough stress periods, but mostly "good" years, we can include this in the model by modifying the constraint to
\[\frac{\sum_j{f_ j}}{T} = \alpha,\]
for some slightly positive $\alpha$, to reflect that the model is calibrated in an economic boom, where observable PDs are low. In this case the model is forced to output higher unconditional PDs than the time mean of the conditional PDs. This parameter $\alpha$ can be a subject of expert judgement or an instrument of regulation. 

\subsubsection{Practical application}
To apply the dynamic model for the conditional (PIT) PD and unconditional (TTC) PD estimation, the following steps are suggested.
\begin{enumerate}
\item Partition the portfolio for which we want to fit the model into disjoint \textit{sub-portfolios} $N = \bigcup_i N_i$ ($i\in I$) such that each one has the same risk profile, meaning that
\begin{itemize}
\item the entire portfolio $N$ is affected by the same systematic factor\footnote{It is not trivial what the factor should include. For example, retail loans might be more exposed to inflation than large firms. However we do not have to know what the factor is, we only assume that it is the same for the entire portfolio.},
\item each $N_i$ has a homogenious entity-specific risk (and approximately the same unconditional PD).
\end{itemize}
\item Specify a (discrete) time frame $t_1\le t_2\le\dots\le t_T$ that spans an entire cycle of the systematic factor that affects the portfolio. The input of the model is a (possibly incomplete) data matrix $d_{i, t}$ for $i\in I$ and $1\le t\le T$, the (empirical) default rate for portfolio $i$ at year $t$
\item Calibrate the model, i.e. define a suitable error function and optimize parameters $p_i, \rho_i \ (i \in I)$ and $f_t (1\le t\le T)$. 
\item Evaluate the model fit, i.e. compare the fitted $p_{i,t}$ values with the empirical $d_{i,t}$ values and assess whether the model is acceptable. Statistically testing the hypothesis $\mathbb{E}(d_{i,t}) = p_{i,t}$ involves a multiple hypothesis testing problem, which requires further investigation.
\item Use the calibrated $p_i$ as TTC PD for sub-portfolio $i\in I$.  
\end{enumerate}

%% file: linear1.tex
Let us start with a special case, when
\begin{itemize}
\item we have a complete data set, meaning that sub-portfolio one-year default rates are observable for all $i$ and $t$ combinations and
\item parameters $\rho_i$ are exogenously known fixed numbers (based on, for example, loan type).
\end{itemize}
By introducing variables $K_i=\Phi^{-1}(p_i)$, $\varphi_{i, t} = \Phi^{-1}(p_{i,t})$ and $y_{i,t}=\varphi_{i,t}\sqrt{1-\rho_i}$ (based on Equations \eqref{eq:PD} and \eqref{eq:yphi}), we get the linear system
\begin{equation}
\begin{aligned}
\label{model:lin_eq}
y_{i, t} &= K_i-\sqrt{\rho_i} f_t\\
\sum f_t &= 0.
\end{aligned}
\end{equation}
By replacing $y_{i, t}$ by its empirical equivalent 
\begin{equation}
\label{eq:eta} 
\eta_{i,t} = \sqrt{1-\rho_i}\Phi\left(d_{i,t}\right)
\end{equation}
where $d_{i, t}$, the default rate of sub-portfolio $i$ in year $t$ is the input of the model, we get the regression model
\begin{equation}
\label{model:lin_reg}
\begin{aligned}
\eta_{i, t} &\approx K_i-\sqrt{\rho_i} f_t\\
\sum f_t &\approx 0
\end{aligned}
\end{equation}
With matrix and vector notations, for $|I| = 3$ sub-portfolios and $T = 4$ years, it is as follows.
\begin{equation}
\label{eq:lin_model}
\begin{bmatrix}
	1 &    &   & -\sqrt{\rho_1}     &                  &                  &                   \\		
	1 &    &   &                    &  -\sqrt{\rho_1}  &                  &                   \\		
	1 &    &   &                    &                  &  -\sqrt{\rho_1}  &                   \\		
	1 &    &   &                    &                  &                  &  -\sqrt{\rho_1}   \\		
	  & 1  &   & -\sqrt{\rho_2}     &                  &                  &                   \\		
	  & 1  &   &                    &  -\sqrt{\rho_2}  &                  &                   \\		
	  & 1  &   &                    &                  &  -\sqrt{\rho_2}  &                   \\		
	  & 1  &   &                    &                  &                  &  -\sqrt{\rho_2}   \\		
	  &   &1   & -\sqrt{\rho_3}     &                  &                  &                   \\		
	  &   &1   &                    &  -\sqrt{\rho_3}  &                  &                   \\		
	  &   &1   &                    &                  &  -\sqrt{\rho_3}  &                   \\		
	  &   &1   &                    &                  &                  &  -\sqrt{\rho_3}   \\		
	0 & 0 &0   & 1                  & 1                & 1                &  1                \\		
\end{bmatrix}\begin{pmatrix}K_1\\K_2\\K_3\\f_1\\f_2\\f_3\\f_4\end{pmatrix} \approx \begin{pmatrix}\eta_{1,1}\\\eta_{1,2}\\\eta_{1,3}\\\eta_{1,4}\\\eta_{2,1}\\\eta_{2,2}\\\eta_{2,3}\\\eta_{2,4}\\\eta_{3,1}\\\eta_{3,2}\\\eta_{3,3}\\\eta_{3,4}\\0\end{pmatrix}
\end{equation}
It is easy to see that the coefficient matrix has full rank, so the system has a unique analytical solution $K_i$ and $f_t$ with the following properties.
\begin{itemize}
\item The last equation is exact, meaning that $\sum f_t = 0$ without error,
\item $K_i = \sum_t \eta_{i,t}\ \forall i\in I$,
\item $K_i = \sum_t y_{i,t} \ \forall i\in I$, where $y_{i,t}$ are the fitted (predicted) value for $\eta_{i,t}$.
\end{itemize}

%% file: linear2.tex
In this section we still assume that parameters $\rho_i$ are exogenous, and discuss what happens in case of missing values. To put it very simply, when input $d_{i,t}$ is incomplete, i.e. it is only available for some $i,t$ pairs, but not for all of them, then some of the rows are not present in Equation \eqref{eq:lin_model}. This can reduce the rank of the coefficient matrix, render the system singular, and ultimately compromise the model. Unfortunately, no simple or easily interpretable characterization is available to determine exactly when this occurs, but we can illustrate it with two examples, as follows.
\begin{itemize}
\item If there is a collection $I_0\subset I$ sub-portfolios and an $S\subset \left\{1...T\right\}$ subset of years, such that $d_{i,t} = \text{\textit{NA}} \quad \forall i \in I_0$ and $t \in S$ and $d_{i,t} = NA$ for all $i \notin I_0$ and $t \notin S$, then the model is unidentifiable.
\item If there exists $i_1,i_2,\dots,i_k\in I$ sequence of sub-portfolios and $1=t_0 \le t_1\le t_2\le\dots\le t_k=T$ sequence of years such that $d_{i_j,t}\neq\text{\textit{NA}} \quad \forall t_{i_j-1}\le t\le t_{i_j}$, then the model is identifiable.
\end{itemize}
Both cases are intuitive. If the bank completely eliminates its entire portfolio and creates a new one from one year to the next, then based on the PIT PDs it is impossible to tell which sub-portfolio has low TTC PD in bad years and which one has high TTC PD in good years. On the other hand, if there is a sequence of overlapping sub-portfolios (each one starts before the next ends) that cover the entire cycle, this is enough to identify the model.

Let us assume that we have an incomplete data set of $d_{i,t}$ default rates (for \textit{some} $(i, t)$ pairs) and transform them to $\eta_{i,t}$ based on Formula \eqref{eq:eta}. If the model is identifiable, then we can calculate $p_i$ and $f_t$ and $p_{i,t}$ based Formula \eqref{eq:pit} for \textit{all} $i$ and $t$ pairs. If $d_{i,t}$ exists, then $p_{i,t}$ is its approximation. However if $d_{i,t}$ is missing, then $p_{i,t}$ is its substitute. To analyze how the TTC PD is related to observable PIT PDs, let $i$ be a portfolio for which $d_{i, t}$ is partially available (it is known for some years but not for the entire time frame), combine Formula \eqref{model:lin_eq} and \eqref{model:lin_reg},
\begin{equation}
\label{model:incomplete_lin_reg}
\eta_{i, t} \approx y_{i, t} = K_i-\sqrt{\rho_i} f_t
\end{equation}
and average both sides of the equation. We get that
\begin{equation}
\label{eq:incomplete_bias}
\frac{\sum_t\eta_{i, t}}{T_i} \approx \frac{\sum_j y_{i, t}}{T_i} = K_i-\sqrt{\rho_i} \frac{\sum_tf_t}{T_i}
\end{equation}
where $T_i$ is the number of years for which $d_{i,t}$ is available. As this is an incomplete average $\sum f_t$ on the right-hand-side is not necessarily zero here, but we can still conclude that
\[\frac{\sum_t\eta_{i, t}}{T_i} \approx \frac{\sum_t y_{i, t}}{T_i} \le K_i \Longleftrightarrow \frac{\sum_tf_t}{T_i} \ge 0.\]
The transformed TTC PD is higher than the mean of the transformed average PIT PD if and only if the mean of the systematic factor over the observed period is higher than its long-time average. For portfolios for which data is only available in a boom period the model fits higher TTC PD than the mean of the observable PIT PDs, and in a recession vice versa.

%% file: nonlinear.tex
Until now, we assumed that correlation parameters $\rho_i$ are fixed constants. In fact, they are not constants, but neither independent parameters to fit. Regulation Basel II specifies a deterministic functional relationship between the TTC PD of a portfolio and its correlation parameter \citep{bcbs2023cre31} based on empirical studies \citep{lopez2004empirical} as follows.
\begin{equation}
\label{eq:rho}
\rho = \rho_{\text{min}}\ \frac{1-e^{-W\cdot PD}}{1-e^{-W}}+\rho_{\text{max}}\left(1-\frac{1-e^{-W\cdot PD}}{1-e^{-W}}\right),
\end{equation}
where the parameters $\rho_{\text{min}}$, $\rho_{\text{max}}$ and $W$ are specified by the regulation and depend on load type. For example in case of corporate loans, $\rho_{\text{min}} = 0.12$, $\rho_{\text{max}} = 0.24$ and $W = 50$, in case of retail exposures they are $0.03$, $0.16$ and $35$, respectively. To include this in the model, all we have to do is to replace $\rho_i$ by $\rho(K_i)$ in Formula \eqref{model:lin_eq} and \eqref{model:lin_reg}, where $\rho:(0, 1)\to\mathbb{R}$ is a function that defines the relationship between the correlation and the TTC PD (the specific function form does not even matter). This change makes the model nonlinear, so we no longer have the theoretical results for the existence and uniqeness of the solution, but in practice it does not change the model dramatically. As function $\rho$ is smooth and monotonic and is not extremely sensitive to its parameter, the nonlinear system is "close to linear", meaning that we can expect it to be efficently solvable by simple iterative methods. We can summarize the steps of the model fitting introduced in Section \ref{section:model} as follows.
\begin{itemize}
\item Consider $d_{i, t}$, the default rate of portfolio $i$ in year $t$ as input, which can be incomplete.
\item For each $i$ and $t$ pairs, for which $d_{i, t}$ is available, let $\varphi_{i, t} = \Phi^{-1}\left(d_{i,t}\right)$
\item Let
\[ K_i, f_t = \argmin\limits_{K_i, f_t} \left\{\left(\varphi_{i,t}\sqrt{1-\rho(K_i)} - \left(K_i-\sqrt{\rho(K_i)} f_t\right)\right)^2\ | \ \sum_t f_t = 0 \right\},\]
where \[\rho(K) = \rho_{\text{min}}\ \frac{1-e^{-W\cdot \Phi(K)}}{1-e^{-W}}+\rho_{\text{max}}\left(1-\frac{1-e^{-W\cdot \Phi(K)}}{1-e^{-W}}\right).\]
\item Output
 \[p_i = \Phi(K_i) \text{ as TTC PD and } p_{i,t} = \Phi\left(\frac{K_i - \sqrt{\rho(K_i)}f_t}{\sqrt{1 - \rho(K_i)}}\right) \text{ as fitted PIT PD.}\]
\end{itemize}

%% file: numeric.tex
\begin{figure}
    \begin{center}
    \includegraphics[width=10cm]{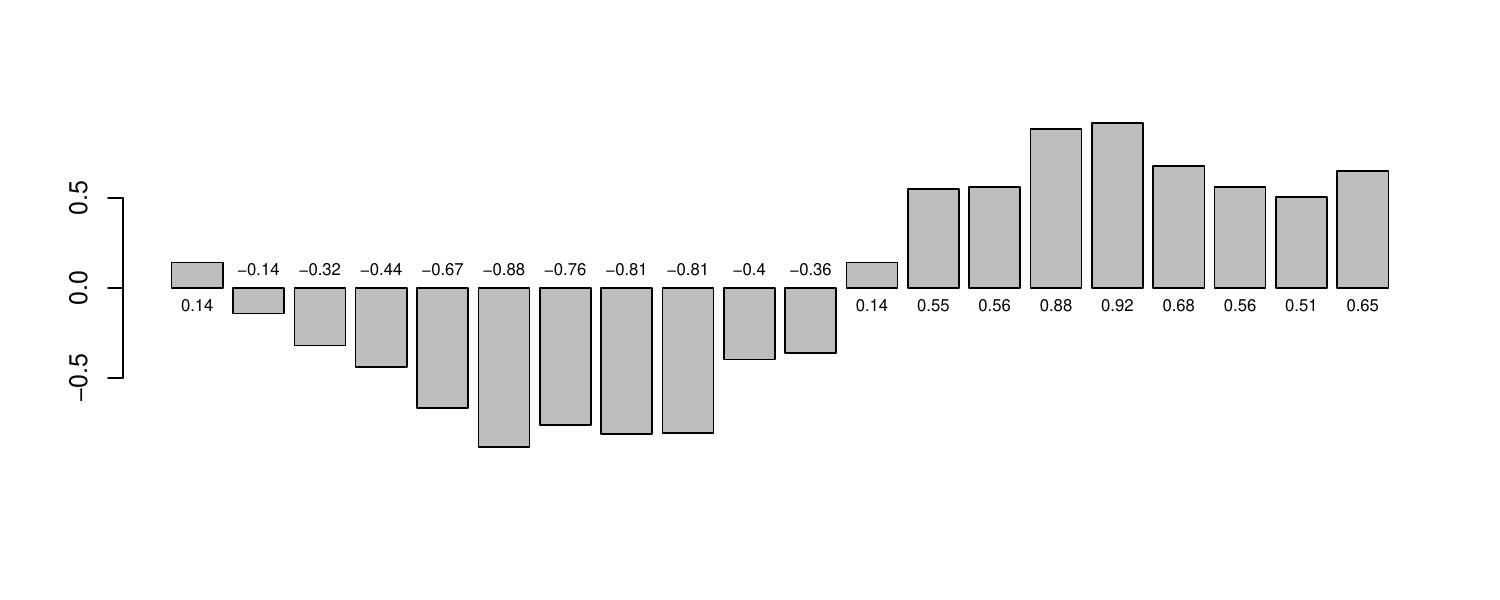}
    \caption{Trajectory of the systematic factor}
    \label{fig:factor}
    \end{center}
\end{figure}

We present an illustration of the above-described methodology on a hypothetical portfolio. In this example the portfolio consists of six sub-portfolios with the theoretical TTC PDs $p_1 = 0.5\%$, $p_2 = 1.7\%$, $p_3 = 3.4\%$, $p_4 = 5.6\%$, $p_5 = 7\%$ and $p_6 = 9\%$ respectively. The evolution of the systematic factor was modeled over a full 20-year cycle, as shown in Figure \ref{fig:factor}, encompassing a recession followed by a boom toward the end of the period. This path represents a selected realization of the standard normally distributed macro factor with first-order autocorrelation.

Default data is randomly generated from a binomial distribution independently for each sub-portfolio $i$ and each year $t$, using the theoretical (true) point-in-time (PIT) PDs calculated from
Equation \eqref{eq:pit}, based on the given through-the-cycle (TTC) PDs and $f$ values. The correlation parameter $\rho$ is derived from Equation \eqref{eq:rho} using the Basel parameters for corporate loans. 

\begin{figure}
    \begin{center}
    \includegraphics[width=\textwidth]{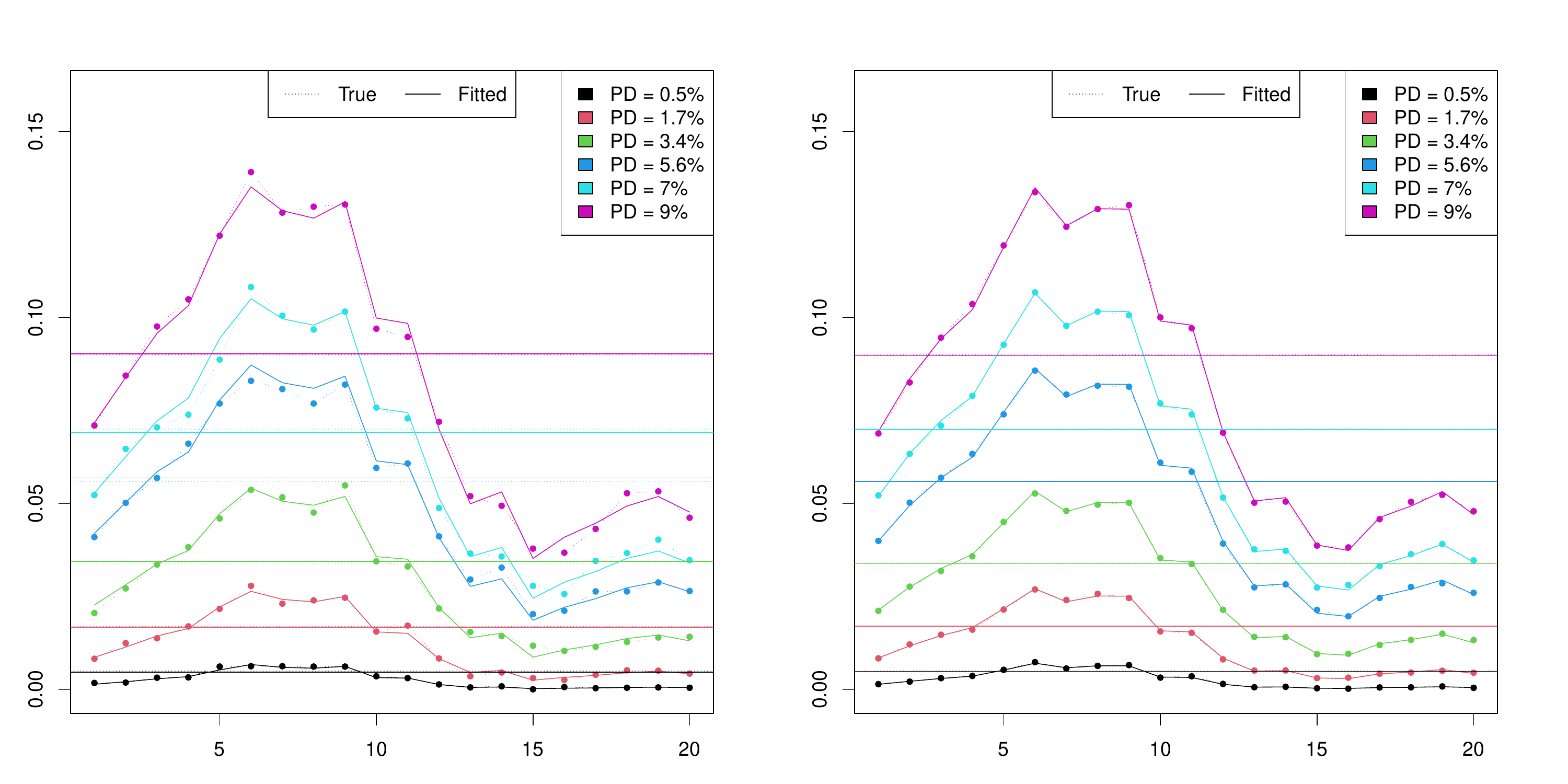}
    \caption{Calibration results for complete data \\The figure shows the theoretical (dotted lines) and fitted (solid lines) values of the conditional and unconditional PDs for six sub-portfolios with varying credit risk levels. In the left chart, each sub-portfolio contains 10\,000 exposures; in the right chart, the sample size per sub-portfolio is 100\,000.}
    \label{fig:output1}
    \end{center}
\end{figure}

The results of the calibration are shown in Figure \ref{fig:output1} for sample sizes $n_1 = 10\,000$ and $n_2 = 100\,000$. The dotted lines represent the theoretical (true) values, while the solid lines show the calibration results. Even for the smaller sample size (left chart), the method provides an almost perfect fit for the unconditional TTC values (straight lines), despite small deviations in the conditional PIT values. A portfolio size of $100\,000$ allows for an excellent fit for both the conditional and unconditional PDs (see right chart). 

\begin{figure}
    \begin{center}
    \includegraphics[width=\textwidth]{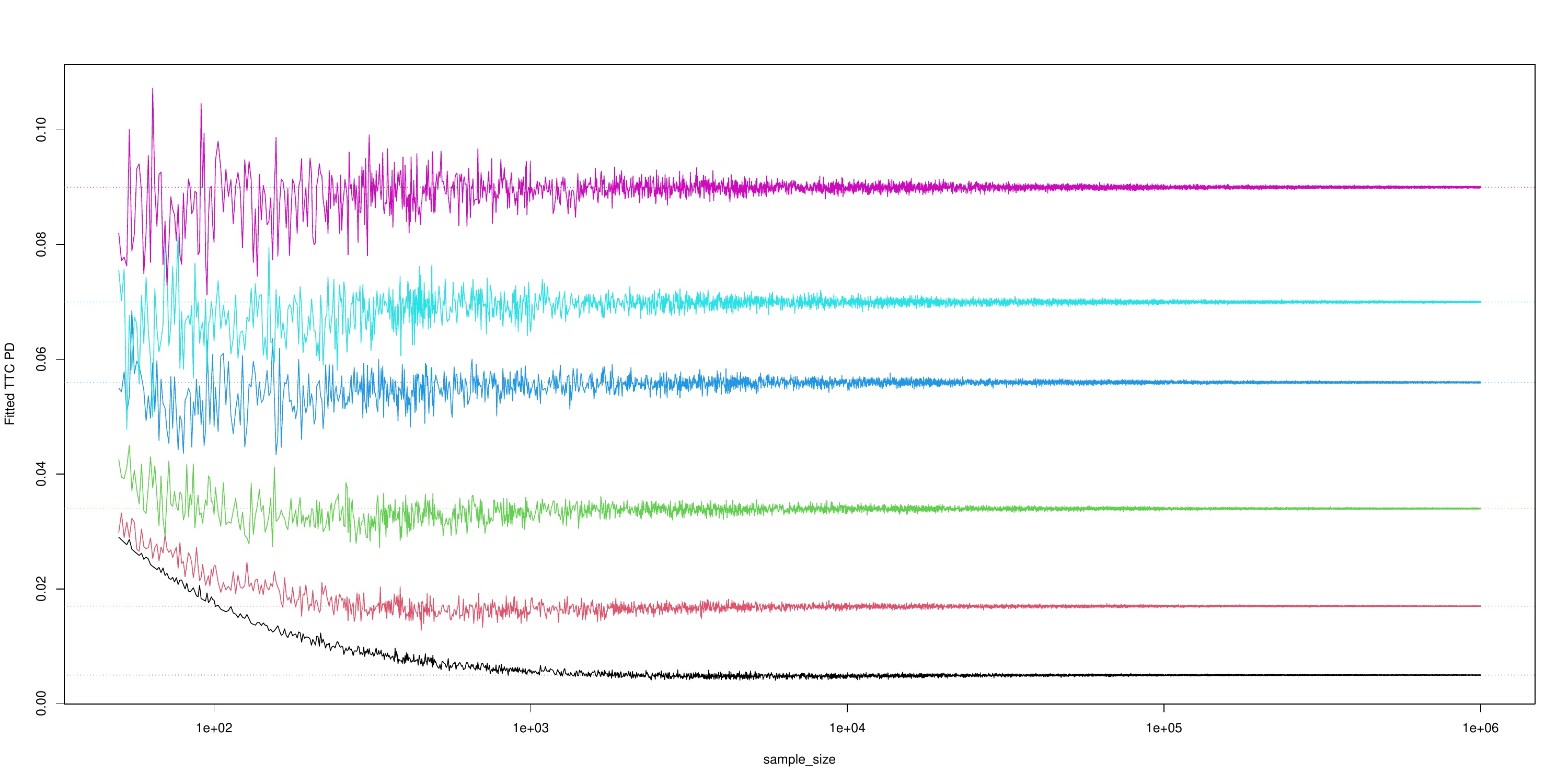}
    \caption{The figure shows the TTC PD estimates as a function of sample size. The straight lines represent the theoretical values to which the estimates converge.}
    \label{fig:output3}
    \end{center}
\end{figure}

The convergence of the model output (fitted TTC PDs) to the theoretical values as a function of sample size is shown in Figure \ref{fig:output3}. Interestingly, the fitted parameters appear to be biased for small samples, which could be a subject of further research. However, for sample sizes of at least a few thousand (which is sufficient for practical applications), the bias vanishes, and for a sample size above $10\,000$, the parameters appear to be asymptotically unbiased and consistent. 

\begin{figure}
    \begin{center}
    \includegraphics[width=\textwidth]{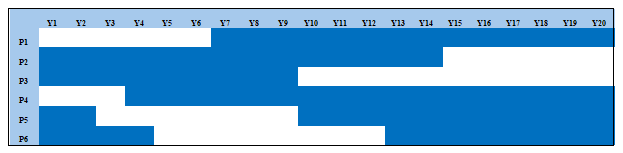}
    \caption{Data availability matrix.\\ Rows represent different sub-portfolios, and columns represent years. Periods with missing values are shown in white. Note that none of the sub-portfolios contain complete data.}
    \label{fig:missing}
    \end{center}
\end{figure}

To test the method on incomplete data, we deleted observations for certain periods across all six sub-portfolios, as shown in Figure \ref{fig:missing}. Periods with available data are highlighted in dark; white cells indicate missing data.

\begin{figure}
    \begin{center}
    \includegraphics[width=\textwidth]{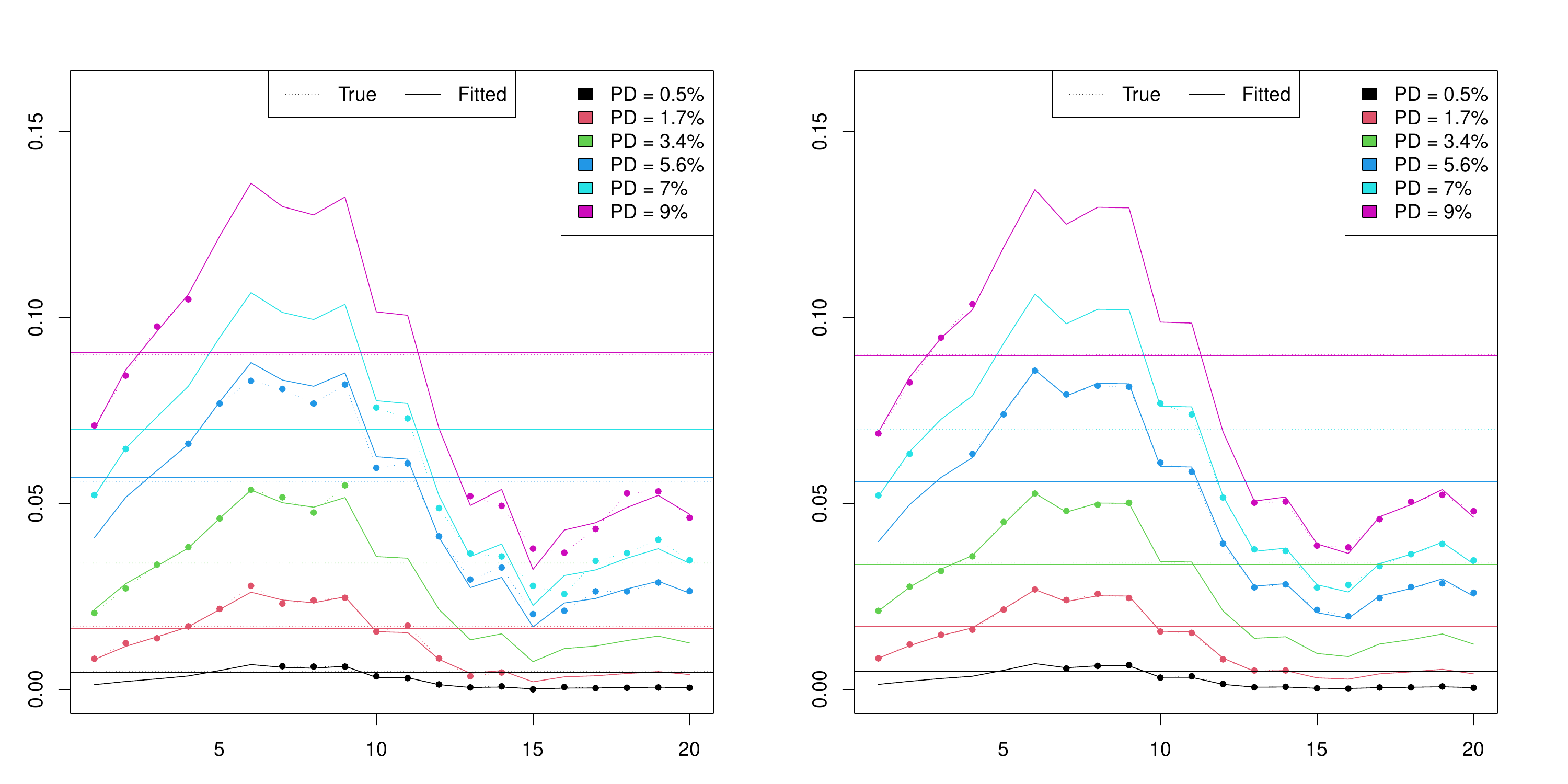}
    \caption{Calibration results for incomplete data \\The figure shows the theoretical (dotted lines) and fitted (solid lines) values of the conditional and unconditional PDs for six sub-portfolios with varying credit risk levels. In the left chart, each sub-portfolio contains 10\,000 exposures; in the right chart, the sample size per sub-portfolio is 100\,000.}
    \label{fig:output2}
    \end{center}
\end{figure}

The calibration process proves to be surprisingly robust -- the model is able to recover its true parameters with accuracy comparable to that achieved with complete data, as shown in Figure 4.5. Although no sub-portfolio contains data for the entire time frame, the model successfully identifies the underlying pattern in the PIT PDs and uses it to infer the missing values. The estimated TTC PDs closely match the time averages of the fitted PIT PDs, which differ markedly from naive estimates based on the incomplete dataset